\documentclass[twocolumn,aps,prb]{revtex4-2}
\usepackage[latin9]{inputenc}
\usepackage{color}
\usepackage{booktabs}
\usepackage{mathtools}
\usepackage{bm}
\usepackage{dsfont}
\usepackage{amsmath}
\usepackage{amssymb}
\usepackage{graphicx}
\PassOptionsToPackage{normalem}{ulem}
\usepackage{ulem}
\usepackage[bookmarks=false,
 breaklinks=false,pdfborder={0 0 1},backref=section,colorlinks=true]
 {hyperref}
\hypersetup{
 pdfborderstyle=}

\makeatletter

\providecommand{\tabularnewline}{\\}

\makeatother

\begin{document}
\title{Plasmon-plasmon interaction in nanoparticle assemblies: role of the
dipole-quadrupole coupling }
\author{Olivier Masset}
 \altaffiliation[]{corresponding author}
\email{olivier.masset@univ-perp.fr}

\author{Roland Bastardis}
\email{roland.bastardis@univ-perp.fr}

\author{Francois Vernay}
\email{francois.vernay@univ-perp.fr}

\address{Laboratoire PROMES CNRS (UPR-8521) \& Universit\'e de Perpignan Via
Domitia, Rambla de la thermodynamique, Tecnosud, F-66100 Perpignan,
FRANCE}
\date{\today}
\begin{abstract}
The synthesis of metallic nanoparticle assemblies is nowadays well-controlled,
such that these systems offer the possibility of controlling light
at a sub-wavelength scale, thanks, in parts to surface plasmons. Determining
the energy dispersion of plasmons likely to couple to light in these
nanostructures is, therefore, a necessary preliminary task on the
way to understanding both their photonic properties and their physical
nature, in particular the role of the quadrupole contribution. Starting
with a general model that takes account of all energy modes, we show
that its low-lying energy dispersion, gained numerically, can be compared
to that of a minimal model that treats dipoles and quadrupoles on
the same footing. The main advantage of the latter relies on the fact
that its formulation is tractable, such that a semi-analytical Bogoliubov
transformation allows one to access the experimentally relevant energy
bands. Based on this semi-analytical derivation, we determine quantitatively
the limit of validity of both the dipole-only model and the presently
proposed dipole and quadrupole model, compared to a full-plasmon-mode
Hamiltonian. The results show that the dispersion relation, which
accounts for dipoles and quadrupoles, is sufficient to capture the
low-energy physics in most experimental situations. Besides, we show
that at small lattice spacing, the contribution of quadrupoles is
dominant around the Brillouin zone center. 
\end{abstract}
\maketitle

\section{Introduction}

Plasmonic materials have proven to show a wide range of properties,
with continuous research ranging from the fields of photonics \citep{Wang},
condensed matter \citep{Grigorenko}, and spectroscopy. As a few striking
examples, quantum entanglement was shown to survive a photon-plasmon-photon
conversion \citep{Altewischer,Fasel}, which justified a quantum description
of the plasmons. A regime of very strong coupling with light was observed
in plasmonic crystals, with the breakdown of the Purcell effect \citep{Mueller}.
The possibility of investigating single-molecule vibrational states
via plasmonic cavities was demonstrated \citep{Chikkaraddy,Arul}.
Among plasmonic materials, ordered plasmonic nanostructures are especially
interesting as the control of their geometry enables the control of
their optical properties \citep{Guo,Schulz}, and could lead to tailored
properties of materials \citep{Basov}. In the theoretical description
of these superlattices, different models have been proposed. Some
models use a classical description of the plasmons via point dipole
\citep{Markel,Evlyukhin1,Zou1} or point multipole models \citep{Swiecicki},
and some use a quantum description of the plasmons via dipolar second
quantization bosonic operators \citep{Weick1,Weick2,Downing,Fernique}
or multi-polar bosonic operators \citep{Bergman3,Manjavacas,Barros,Finazzi}.
The second quantized formalism becomes quite handy in describing the
coupling with other fundamentally quantum systems, like molecules
\cite{Neuman} or quantum dots \cite{Gupta} which makes it the formalism
of choice here.

Regardless of the model chosen, it is always desirable to have a minimal
model for the description of the physics at hand. That is why limiting
the description to a coupled dipole model is often the chosen option.
However, it is well established that this model stands for particles
distant enough from one another. Indeed, a dipole-only model fails
to describe the physics when the center-to-center distances ($d$)
gets smaller than three times the radius ($d/R<3$) of the nanoparticles
\citep{Park}. This is due to plasmon-plasmon induced hybridization
\citep{Prodan}, and it is observed even for small particles described
in the quasistatic limit. This fact is particularly acute in recent
interesting experimental realizations of colloidal crystals and other
nanoparticle assemblies \citep{Mueller,Arul}, for which the nanoparticle
center-to-center distance is well below the dipole model limit. Besides,
considering coupling to fast electrons and their near fields requires
higher order plasmons even for small particles, down to 4nm \citep{Raza}.
All these facts combine to suggest that a simple analytical expression
of the low-energy plasmonic bands is necessary to help interpret future
experimental data.

In the present paper, we intend to take stock of this knowledge by
building a minimal semi-analytic description, in line with earlier
investigations. This description will necessarily go beyond a simple
dipole model, as we have previously emphasized. We thus investigate
the importance of higher order plasmonic modes in dense assemblies.
For the sake of clarity and simplicity, we choose a 1D chain of spherical
nanoparticles as a model system, and we propose a minimal Hamiltonian
leading to analytic expressions for the lowest energy plasmonic modes.
Specifically, we discuss our derivation in light of previous similar
approaches and show that taking account of the quadrupoles is sufficient,
as it accurately describes the physics relevant to experiments. The
present minimal model implies tractable Hamiltonian matrices, and
therefore combines two advantages: it allows one to derive a semi-analytical
expression of the low-lying excitation dispersion, and the contribution
of dipoles vs quadrupoles can be easily investigated. Besides, it
sets the stage for a future coupling to photons, which is more straightforward
in contrast to full multipole models.

The paper is organized as follows: In the first section, we present
the model as a truncation of the multipole model described in Refs.\citep{Park,Bergman1},
and show that it is comparable to Barros \emph{et al.} approximation
\citep{Barros}. By inspecting the physical nature of the terms in
the expression of the dispersion, we briefly argue why it is not possible
to simplify this model further. Then, we present the plasmonic band
structure obtained through this method and comment on the limit of
validity. The paper closes with our conclusion and perspectives.

\section{Model}

The optical properties of nanoparticle assemblies are formally described
by the following generic Hamiltonian 
\begin{equation}
\mathcal{H}=H_{pl}+H_{\gamma}+H_{pl-\gamma}
\end{equation}
where $H_{pl}$ stands for the plasmonic degrees of freedom, whereas
$H_{\gamma}$ and $H_{pl-\gamma}$ respectively describe the photonic
bath and its coupling to the plasmons. For the sake of clarity, this
paper focuses solely on determining the minimal $H_{pl}$ relevant
to experiments. Indeed, the derivation of reliable physical observables
first implies correctly describing the plasmonic Hamiltonian. 
\begin{figure}
\centering{}\includegraphics[width=1\columnwidth]{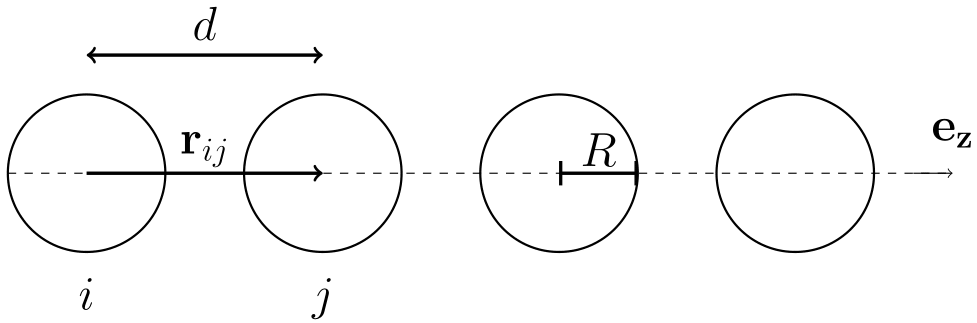}
\caption{Schematic representation of a 1D chain of spherical particles of radius
$R$ with center-to-center distance $d$. The chain runs along the
$z$ axis.}
\label{fig:chain} 
\end{figure}

In the present work, we consider an assembly of metallic spherical
nanoparticles embedded in a dielectric matrix as depicted, in Fig.\ref{fig:chain},
for a 1D case. In this system, each nanoparticle hosts a discrete
spectrum of localized surface plasmons $\left|\phi_{\ell,m}\right\rangle $
with corresponding energies $\hbar\omega_{\ell}$, where $\ell$ stands
for the order of the mode and $m$ stands for its azimuthal index.
In second quantization, one can thus write a Hamiltonian $\sum_{\ell,m}\hbar\omega_{\ell}b_{\ell m}^{\dagger}b_{\ell m}$,
where $b_{\ell m}^{\dagger}$ (respectively $b_{\ell m}$) creates
(annihilates) a localized surface plasmon. For spherical particles,
in the quasistatic limit, one has $\omega_{\ell}=\sqrt{\frac{\ell}{2\ell+1}}\omega_{p}$,
where $\omega_{p}$ stands for the plasma frequency
of the metal. To give an order of magnitude, from
experimental values, the bulk plasmon frequency of silver is $\hbar\omega_{p}^{{\rm Ag}}=9.04{\rm \ eV}$ and
the one of gold is $\hbar\omega_{p}^{{\rm Au}}=8.89{\rm \ eV}$ \citep{Zeman}.
These relatively high-energy values translate to the important role
played in plasmonics by the nanoparticles' size, as well as the surrounding
matrix in assemblies \citep{Miwa}.

Within an assembly, the plasmon modes interact with one another, implying
a hybridization, thereby forming bands. The plasmon-plasmon interaction
terms can be evaluated either from overlap integrals $Q_{\ell m\ell'm'}$
\citep{Bergman1,Bergman2,Park} or from the Coulomb interaction in
hydrodynamics models \citep{Prodan}. Taking this interaction into
account and using the overlap formulation the plasmonic Hamiltonian
reads 
\begin{widetext}
\begin{equation}
\begin{array}{ccc}
H_{pl} & = & {\displaystyle \sum_{\mathbf{k}}\sum_{\ell,m}}\hbar\omega_{\ell}b_{\mathbf{k}\ell m}^{\dagger}b_{\mathbf{k}\ell m}+{\displaystyle \sum_{\mathbf{k}}\frac{\hbar\omega_{p}^{2}}{4}\sum_{\ell,m}\sum_{\ell',m'}\frac{Q_{\mathbf{k}\ell m\ell'm'}}{\sqrt{\omega_{\ell}\omega_{\ell'}}}}\left[b_{\mathbf{-k}\ell m}\left(b_{\mathbf{k}\ell'm'}+b_{\mathbf{-k}\ell'm'}^{\dagger}\right)+b_{\mathbf{k}\ell m}^{\dagger}\left(b_{\mathbf{k}\ell'm'}+b_{\mathbf{-k}\ell'm'}^{\dagger}\right)\right]\end{array},\label{eqn:Hpl}
\end{equation}
with the overlap for a 1D-chain given below, its general expression
for other geometries can be found in Ref. \citep{Park} 
\begin{align}
\begin{array}{lll}
Q_{\mathbf{k}\ell m\ell'm'} & = & \left(\frac{R}{d}\right)^{\ell+\ell'+1}\left(-1\right)^{\ell'+m}\sqrt{\frac{\ell\ell'}{(2\ell+1)(2\ell'+1)}}\frac{(\ell+\ell')!}{\sqrt{(\ell+m)!(\ell'+m)!(\ell-m)!(\ell'-m)!}}\\
 &  & \times\left[\text{Li}_{\ell+\ell'+1}\left(e^{ikd}\right)+\left(-1\right){}^{\ell+\ell'}\text{Li}_{\ell+\ell'+1}\left(e^{-ikd}\right)\right]\delta_{mm'}
\end{array}\label{eqn:Qlmk}
\end{align}
and the Polylog function 
\begin{equation}
\text{Li}_{n}(z)=\sum_{k=0}^{\infty}\frac{z^{k}}{k^{n}}
\end{equation}
\end{widetext}

The main advantage of this formulation relies on the fact that, in
its general expression, the Eq. (\ref{eqn:Hpl}) allows one to describe
any plasmonic nanoparticle assembly within the quasistatic limit,
provided that the corresponding overlap term $Q_{\bm{k},\ell m\ell'm'}$
is correctly evaluated. To avoid unnecessary cumbersome expressions
that would obscure the discussion, the present work limits its quantitative
investigation to 1D systems. Nevertheless, 2D and
3D situations can be addressed within the same framework. Yet, care
has to be taken while computing the overlap of Eq. (\ref{eqn:Qlmk})
from the real space expression in \citep{Park}.

Owing to the symmetry of the chain, the present hamiltonian is block
diagonal in $m$, and modes with different azimuthal indexes do not
interact. In principle, as shown by Park and Stroud \citep{Park}
a sharp description of the whole physics at play requires a large
number of modes. However, if one focuses on the low-energy part of
the spectrum, truncating at $\ell_{max}=2$ is sufficient, as shall
be checked in Section \ref{subsec:Superlattice-spacing}. In addition,
from an experimental point of view this approximation appears reasonable:
For instance, the plasmonic modes that will easily couple to light
are $\ell=1$-dipolar modes and, as we shall see, those are strongly
influenced by $\ell=2$-quadrupolar modes. Limiting oneself to $\ell_{max}=1$
in Eq. (\ref{eqn:Hpl}) leads to 
\begin{align}
H_{dip}= & \sum_{\mathbf{k}}\sum_{m}\left\{ \hbar\omega_{1}b_{\mathbf{k}1m}^{\dagger}b_{\mathbf{k}1m}\right.\nonumber \\
 & \left.+\frac{\hbar\omega_{p}^{2}}{2\omega_{1}}Q_{\mathbf{k}11m}\left[b_{\mathbf{-k}1m}\left(b_{\mathbf{k}1m}+b_{\mathbf{-k}1m}^{\dagger}\right)+{\rm h.c.}\right]\right\} 
\end{align}
This latter Hamiltonian restricted to dipolar modes is investigated
by Allard and Weick in Ref.\citep{Weick1}. In the same spirit, but
keeping in addition quadrupolar modes, we shall truncate the Hamiltonian
in Eq. (\ref{eqn:Hpl}) at order $\ell_{max}=2$. This approximation
corresponds to that of Ref.\citep{Barros}.

Yet, the derivation of the Hamiltonian by Barros \emph{et al.} is
slightly different from the present one. It provides expressions of
canonically conjugate momenta for quadrupolar excitations allowing
for the coupling with a photonic bath, in the form of the quadrupolar
tensor, and the expression is more involved. Nevertheless, in the
quasistatic limit, the two models come together and provide access
to the same physics. Besides, the present formulation allows one to investigate the role of higher-order
multipoles, when Ref. \citep{Barros}, is limited to quadrupoles due
once again to the difficulty of working out the coupling to free space
electromagnetic field.

The diagonalization of dipole+quadrupole Hamiltonian resulting from
the $\ell_{max}=2$-truncation of Eq. (\ref{eqn:Hpl}) is then gained
by performing a Bogoliubov transformation. This unitary transformation
is obtained once one has introduced the operators $\mu_{\mathbf{k}\ell}^{\dagger}$
and $\mu_{\mathbf{k}\ell}$ as linear combination of the original
annihilation and creation operators $b$ and $b^{\dagger}$. Such
that 
\begin{align}
\mu_{\mathbf{k}\ell}=u_{\mathbf{k}\ell}b_{\mathbf{k}\ell}+v_{\mathbf{k\ell}}b_{\mathbf{-k}\ell}^{\dagger}+\sum_{\ell'\neq\ell}\left(w_{\mathbf{k}\ell'}b_{\mathbf{k}\ell'}+x_{\mathbf{k}\ell'}b_{\mathbf{-k}\ell'}^{\dagger}\right)\label{eqn:bogoliubov}
\end{align}
where $u_{\mathbf{{k\ell}}},$ $v_{\mathbf{{k\ell}}},$
$w_{\mathbf{{k\ell'}}}$ and $x_{\mathbf{{k\ell'}}}$ represent a
set of complex coefficients which are nothing but the elements of
the unitary matrix involved in the Bogoliubov transformation. Their
expression do not enter the computation of the energies conducted
below from the commutators expressions. This expression
with $\left(\ell,\ell'\right)=\left(1,2\right)$, enables to rewrite
the Hamiltonian as 
\begin{equation}
H_{pl}=\sum_{\mathbf{k}\in\mathcal{D}}\sum_{\ell}\hbar\omega_{\mathbf{k}\ell}\mu_{\mathbf{k}\ell}^{\dagger}\mu_{\mathbf{k}\ell}.
\end{equation}
The sum in the reciprocal space runs over the half-space $\mathcal{D}$.
Indeed, one has to pay attention to the fact that in the definition
of Eq. (\ref{eqn:bogoliubov}) the original bosonic operators carry
either a positive or a negative $k$. The expressions for the commutator
$[\mu_{\mathbf{k}\ell},H_{pl}]$ lead to solving $\det(M-\omega_{k}\mathds{1})=0$
with 
\begin{equation}
M=\begin{pmatrix}\omega_{1}+f_{\mathbf{k}11m} & -f_{\mathbf{k}11m} & f_{\mathbf{k}12m}^{*} & f_{\mathbf{k}12m}\\
-f_{\mathbf{k}11m} & \omega_{1}+f_{\mathbf{k}11m} & f_{\mathbf{k}12m} & f_{\mathbf{k}12m}^{*}\\
f_{\mathbf{k}12m} & f_{\mathbf{k}12m}^{*} & \omega_{2}+f_{\mathbf{k}22m} & -f_{\mathbf{k}22m}\\
f_{\mathbf{k}12m}^{*} & f_{\mathbf{k}12m} & -f_{\mathbf{k}22m} & \omega_{2}+f_{\mathbf{k}22m}
\end{pmatrix}
\end{equation}
where $f_{\mathbf{k}\ell\ell'm}$ is introduced 
\begin{equation}
f_{\mathbf{k}\ell\ell'm}=\frac{\omega_{p}^{2}}{4}\frac{Q_{\mathbf{k}\ell\ell'm}+Q_{\mathbf{-k}\ell'\ell m}}{\sqrt{\omega_{\ell}\omega_{\ell'}}}
\end{equation}
The low-lying energy dispersion across the Brillouin zone $\hbar\omega_{k}$
is given by

\begin{align}
\omega_{k}^{2} & =\frac{1}{2}\left(\omega_{1}^{2}+\omega_{p}^{2}Q_{\mathbf{k}11m}+\omega_{2}^{2}+\omega_{p}^{2}Q_{\mathbf{k}22m}\right)\nonumber \\
 & -\frac{1}{2}\sqrt{\left(\omega_{1}^{2}+\omega_{p}^{2}Q_{\mathbf{k}11m}-\omega_{2}^{2}-\omega_{p}^{2}Q_{\mathbf{k}22m}\right)^{2}+4\omega_{p}^{4}|Q_{\mathbf{k}12m}|^{2}}\label{eqn:wk}
\end{align}

The inspection of Eqs. (\ref{eqn:Qlmk}) and (\ref{eqn:wk}) shows
that the energy contains: dipole-dipole interaction terms $Q_{11}\propto\left(R/d\right)^{3}$,
dipole-quadrupole interaction terms $Q_{12}\propto\left(R/d\right)^{4}$,
and quadrupole-quadrupole interaction terms $Q_{22}\propto\left(R/d\right)^{5}$,
each of which having a non-trivial dependence in $k$.

\begin{figure*}
\begin{centering}
\includegraphics[width=1\textwidth]{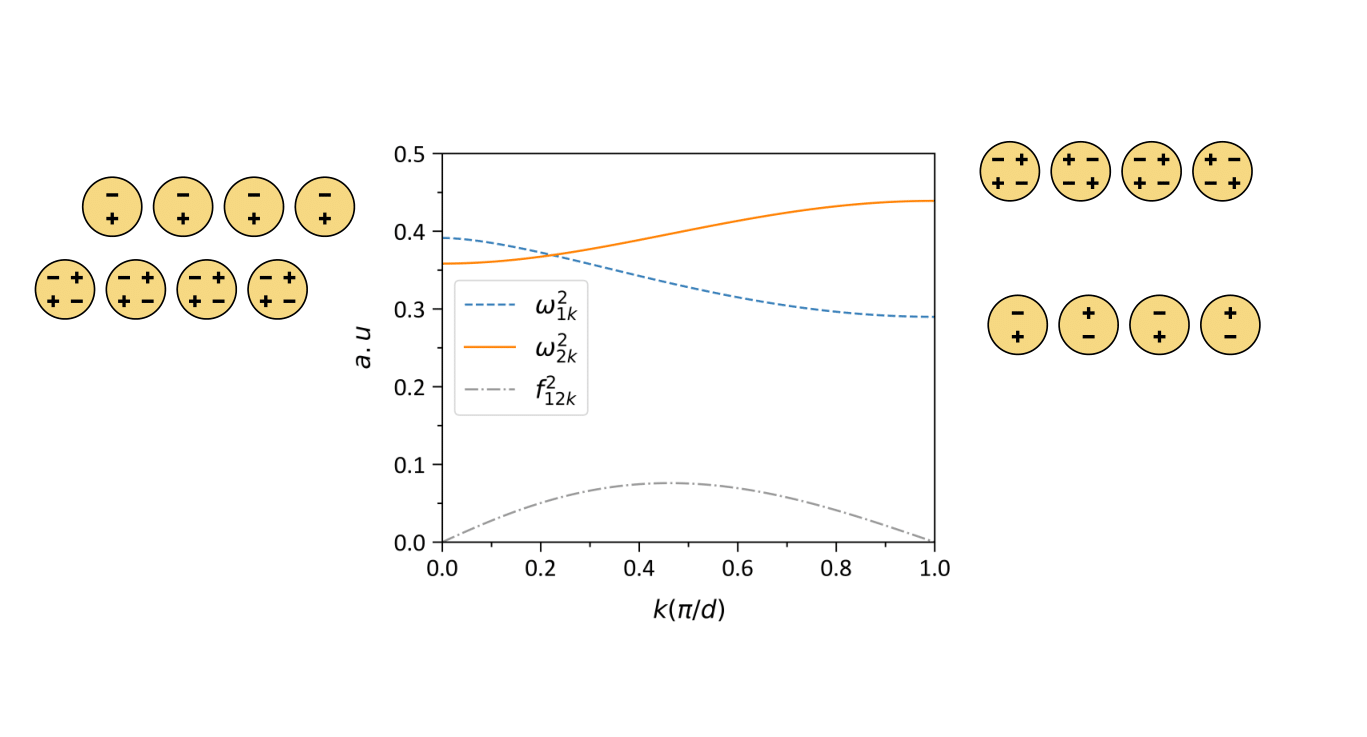} 
\par\end{centering}
\caption{Terms appearing in the dispersion relation of the model with schematic
descriptions of the surface charge densities associated to the $k=0$
and $k=\pi/d$ modes. The dotted blue line corresponds to interacting
dipoles, while the orange solid line corresponds to a chain of interacting
quadrupoles. As pointed out in the text, the bands cross for dense
enough chains, here the values correspond to $d/R=2.4$ and $m=1$. }
\label{fig:omega2} 
\end{figure*}

Two interacting subspaces enter Eq. (\ref{eqn:wk}): that of dipoles
and that of quadrupoles. This observation leads to the rewriting of
the dispersion as $\omega_{k}^{2}=\frac{1}{2}(\omega_{1k}^{2}+\omega_{2k}^{2})-\frac{1}{2}\sqrt{\left(\omega_{1k}^{2}-\omega_{2k}^{2}\right)^{2}+4f_{12k}^{2}}$.
Besides, the analysis of the amplitudes of the respective elements
$\omega_{1k}$, $\omega_{2k}$ and $f_{12k}$ over the Brillouin zone
shows that $f_{12k}^{2}\ll\omega_{1k}^{2},\omega_{2k}^{2}$ as can
be seen in Fig. \ref{fig:omega2}. Therefore, it is tempting to perform
a perturbative expansion of Eq. (\ref{eqn:wk}) with respect to the
parameter $\eta$

\begin{equation}
\eta(k)=\frac{|Q_{\mathbf{k}12m}|}{|\omega_{1}^{2}+\omega_{p}^{2}Q_{\mathbf{k}11m}-\omega_{2}^{2}-\omega_{p}^{2}Q_{\mathbf{k}22m}|}\label{eqn:eta}
\end{equation}

This would correspond to a downfolding of the two subspaces previously
mentioned to an effective one corresponding to dipoles dressed by
quadrupolar excitations. However, one has to notice in Fig. \ref{fig:omega2}
that for dense chains with $d/R<3$, the dipolar band $\omega_{1k}$
and the quadrupolar band $\omega_{2k}$ cross as one goes from $k=0$
to $\pi/d$. Consequently $\eta$ diverges at this crossing and the
nature of the low-energy mode changes. All these facts together suggest
that the minimal model required to describe the low-energy excitations
in an assembly of plasmonic nanoparticles is given by the Hamiltonian
of Eq. (\ref{eqn:Hpl}) with $\ell_{max}=2$. To support these words,
we use the dispersion relation (\ref{eqn:wk}) for different chain
densities and different plasmon polarizations (\textit{i.e.,} different
$m$) and compare it to other models. In the following section we
consider chain densities relevant to experiments and outside of the
dipole model validity domain.

\section{Results and discussion}

\subsection{Superlattice spacing $d$}

\label{subsec:Superlattice-spacing}

Experimentally, dense metallic nanoparticle assemblies can be synthesized
as colloidal crystals formed with small ligands \citep{Mueller},
with inter-particle distances down to $1\sim4\ {\rm nm}$. There has
been remarkable success in controlling the lattice parameter for tunable
crystals with DNA strands \citep{Macfarlane} or polymer shells \citep{Karg}.
These experimental realizations generally give rise to situations
for which $2.4\lesssim d/R\lesssim3$. Denser assemblies with $1\ {\rm nm}$
particle separations or less are also of experimental relevance. However,
in that case, tunneling effects occur \citep{Scholl2}, and one has
to resort to a full quantum description, which is beyond the scope
of the present work. Having that in mind, the quantitative analysis
described below involves three distances: $d/R=3$, $d/R=2.6$ and
$d/R=2.4$.

The plasmon-plasmon interaction part of the Hamiltonian in Eq. (\ref{eqn:Hpl})
contains a prefactor, explicitly written in Eq. (\ref{eqn:Qlmk}),
with an algebraic decay $d^{-(\ell+\ell'+1)}$. This describes how
the superlattice spacing influences the relative importance of the
plasmonic multipoles, with higher-order modes gaining more importance
as the particles get closer together. Nevertheless, the experimentally
relevant modes are those likely to couple to light, namely and principally
dipolar ones. Therefore, it becomes essential to understand qualitatively
and quantitatively the role of these higher-order modes on the lower
energy plasmonic band, which contains dipolar modes.

\begin{figure}
\centering{}\includegraphics[width=0.5\textwidth]{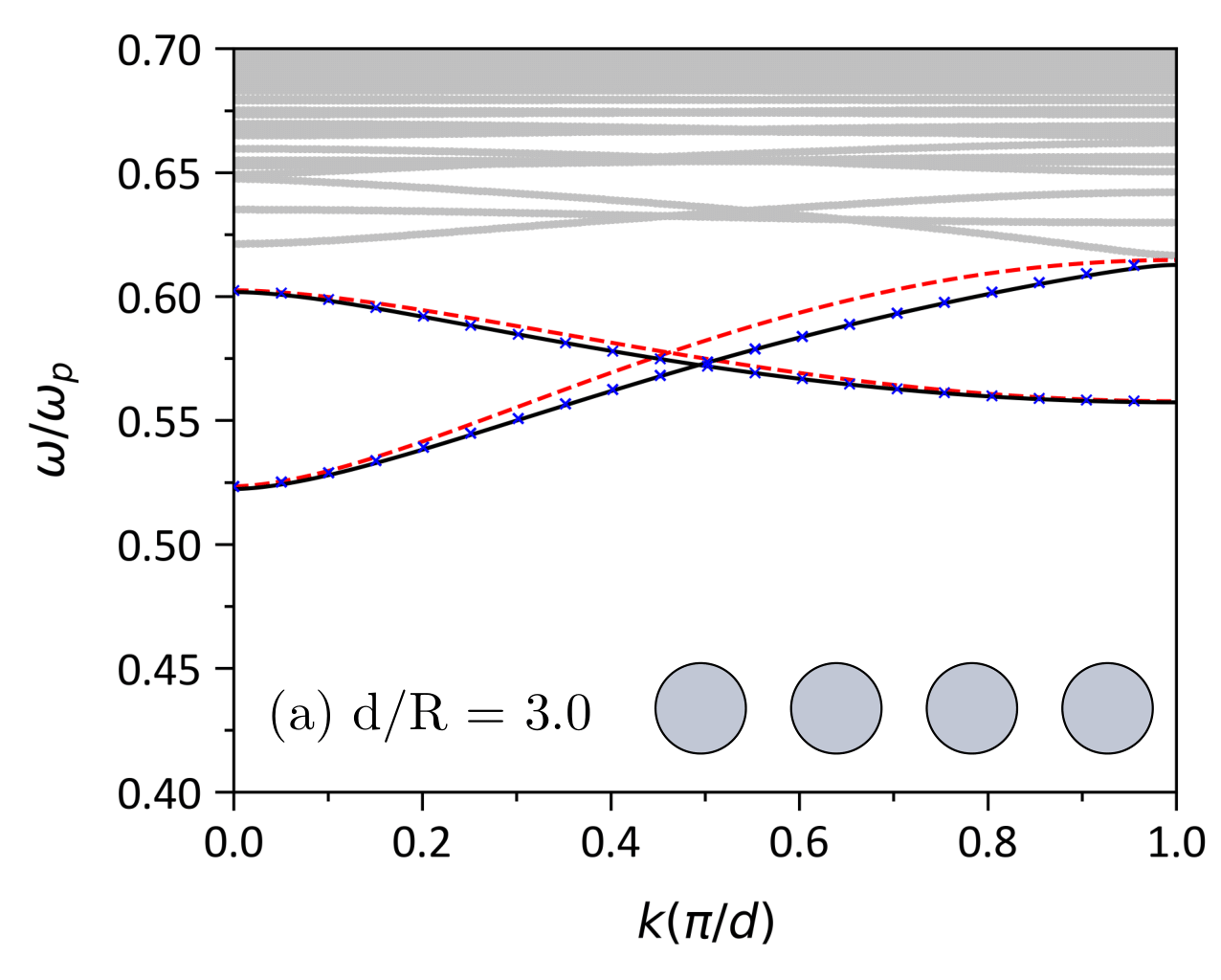}
\includegraphics[width=0.5\textwidth]{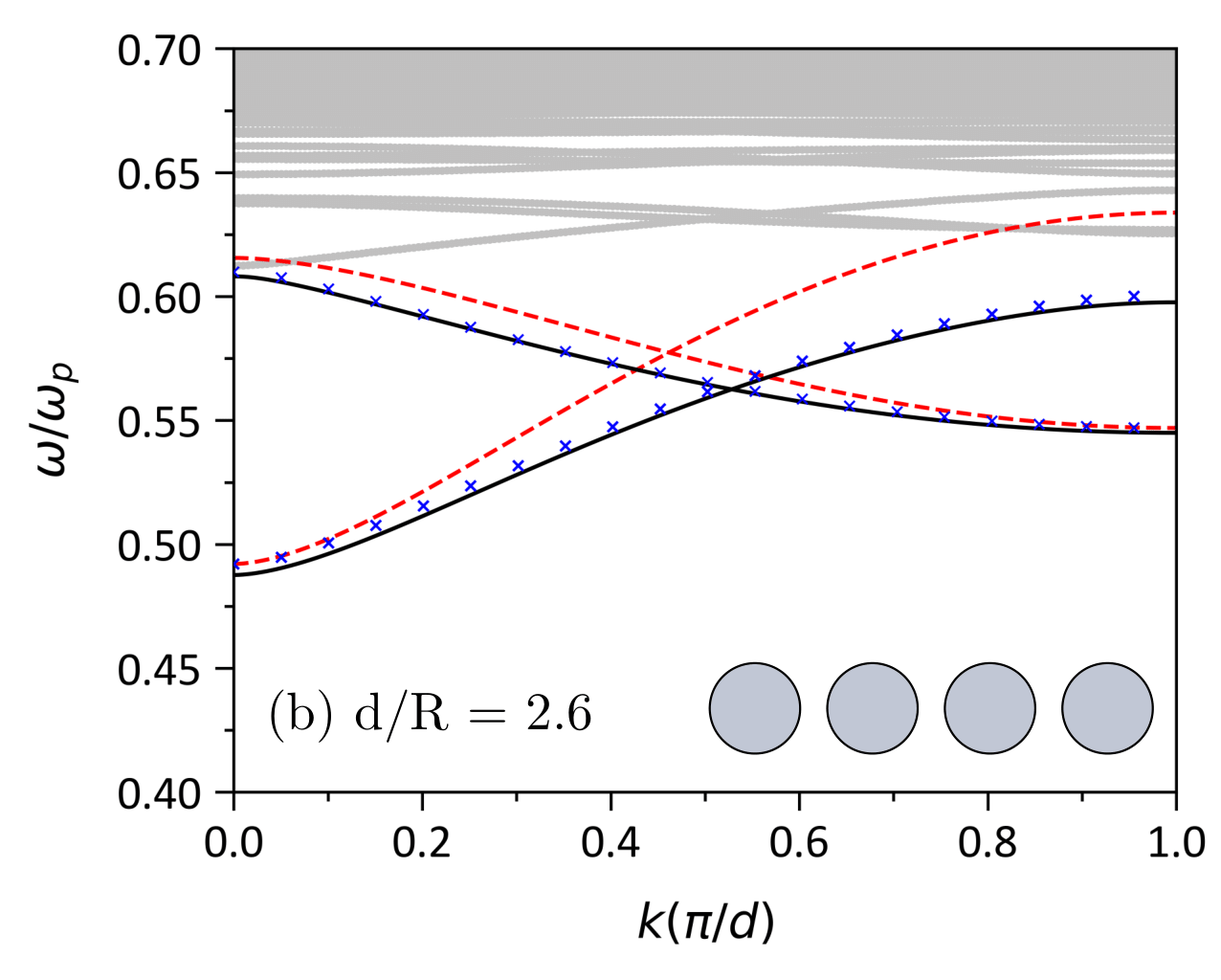} \includegraphics[width=0.5\textwidth]{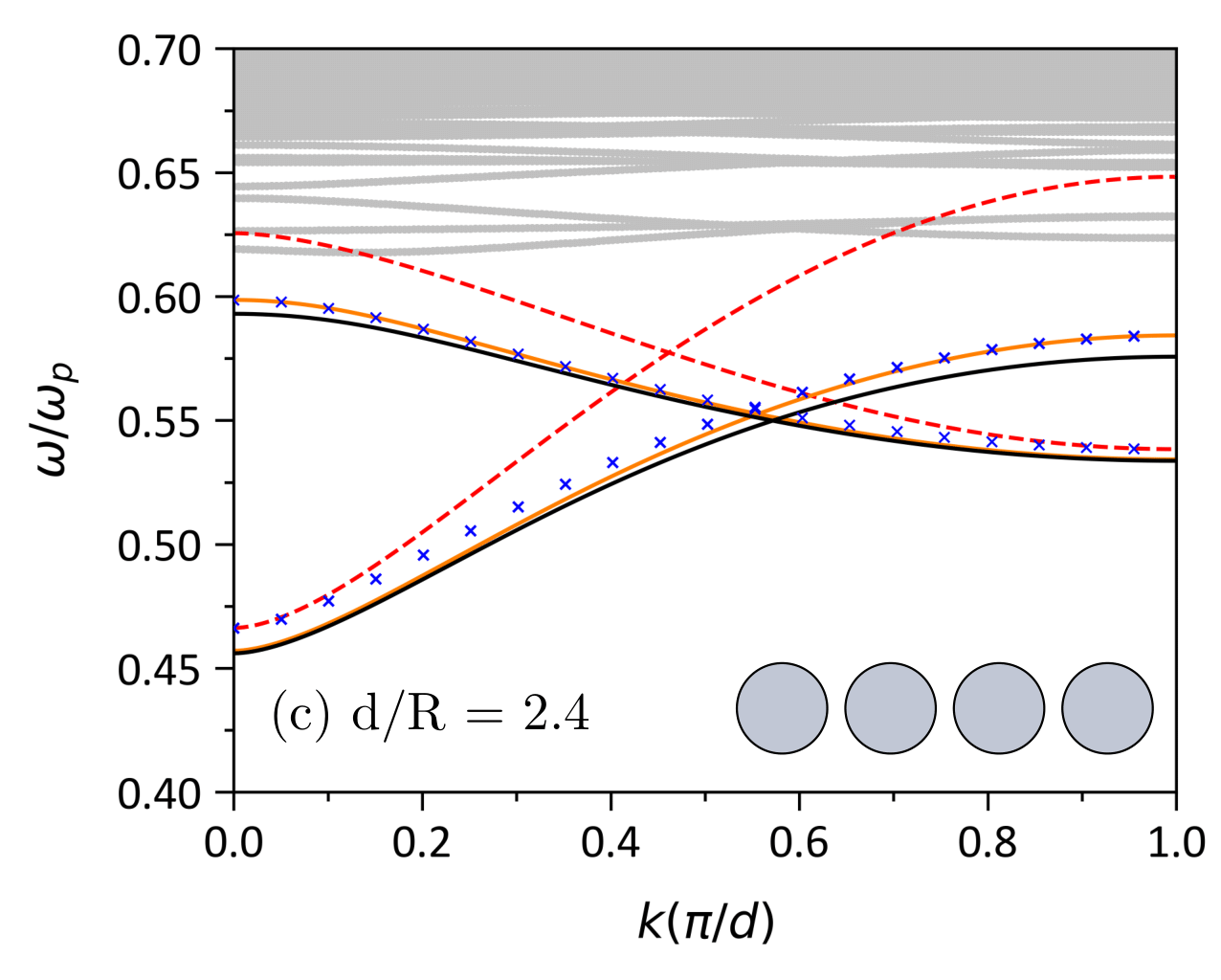}
\caption{Plasmonic band diagram of a chain of spherical metallic nanoparticles
of radius $R=10$ nm with three different lattice spacing (a) $d=30\ {\rm nm}$,
(b) $d=26\ {\rm nm}$ and (c) $d=24\ {\rm nm}$. Black lines are calculated
with $\ell_{max}=20$, red dashed lines are the dipole model eigenvalues
with $\ell_{max}=1$, and the blue crosses are the $\ell_{max}=2$
model taking into account quadrupoles. The orange
solid lines in (c) correspond to $\ell_{max}=3$ for comparison.
In (a), (b) and (c) the high energy bands calculated with $\ell_{max}=20$
are grayed out for readability. Insets show the corresponding chain
densities.}
\label{fig:bands} 
\end{figure}

The dispersion of the different modes across the Brillouin zone is
shown in Fig. \ref{fig:bands} for each case: $d/R=3$, $d/R=2.6$,
and $d/R=2.4$. For the sake of simplicity, one can
take a formal example of $R=10{\rm \ nm}$, which is relevant to the
quasistatic limit and provides interparticle distances of 24 nm in
the $d/R=2.4$ case, in line with some typical colloidal crystals
\citep{Macfarlane}. In fact, as the model is scale
invariant, its limitation depends mainly on the quasistatic approximation
validity. The converged $\ell_{max}=20$-model is plotted as gray
and solid black lines, whilst the red dashed lines correspond to the
dipole-model, and the blue crosses to the dipole+quadrupole model.
The two low-lying highlighted bands are directly linked to dipolar
modes. This is obvious if one looks at the $d/R=3$ case, where the
three models produce nearly the same dispersion, and the deviation
from a pure dipole model gets more pregnant as the inter-particle
distance gets smaller. To understand the underlying mechanism, it
is necessary to determine the nature of each band.

\begin{figure}
\centering{}\includegraphics[width=0.5\textwidth]{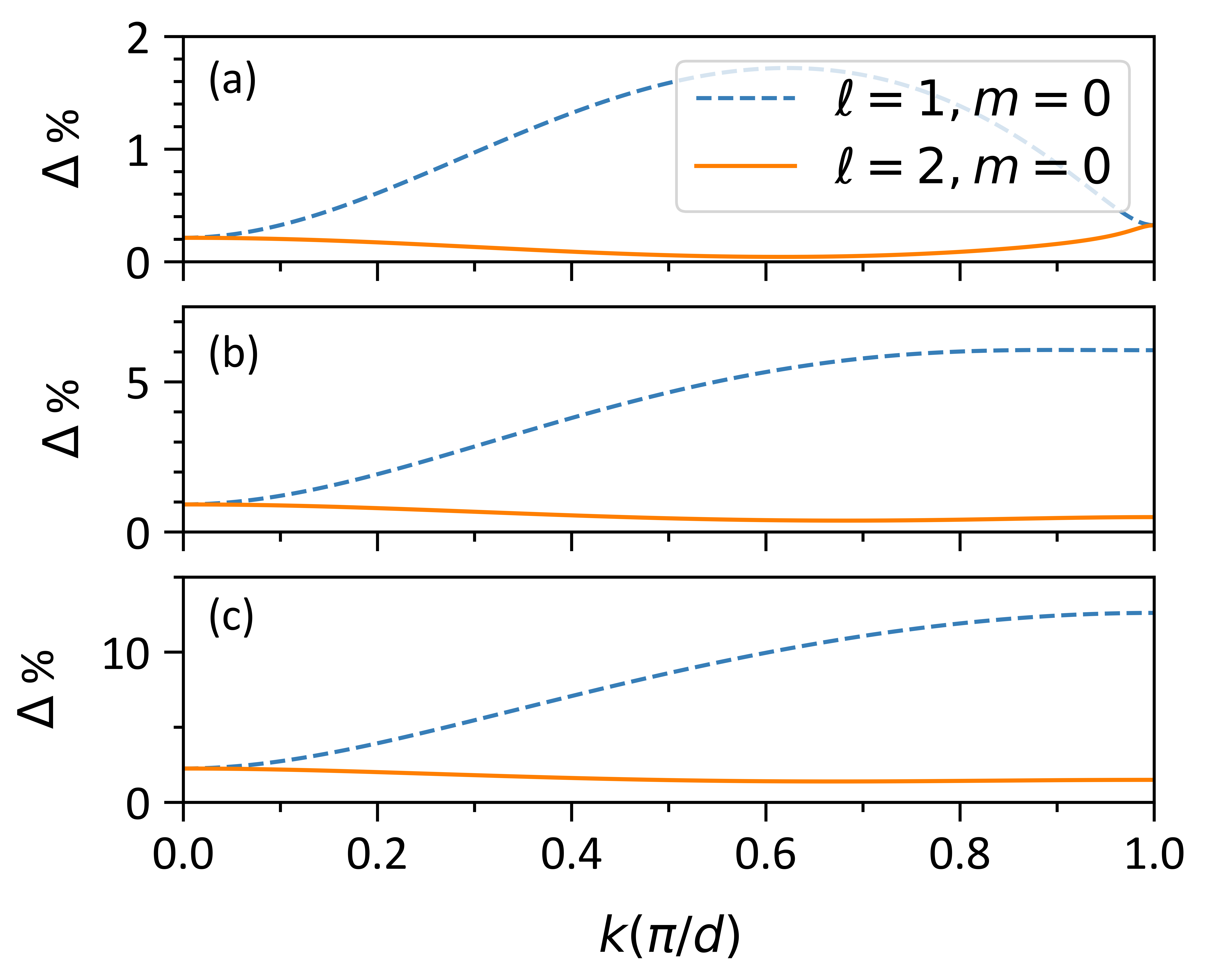}
\includegraphics[width=0.5\textwidth]{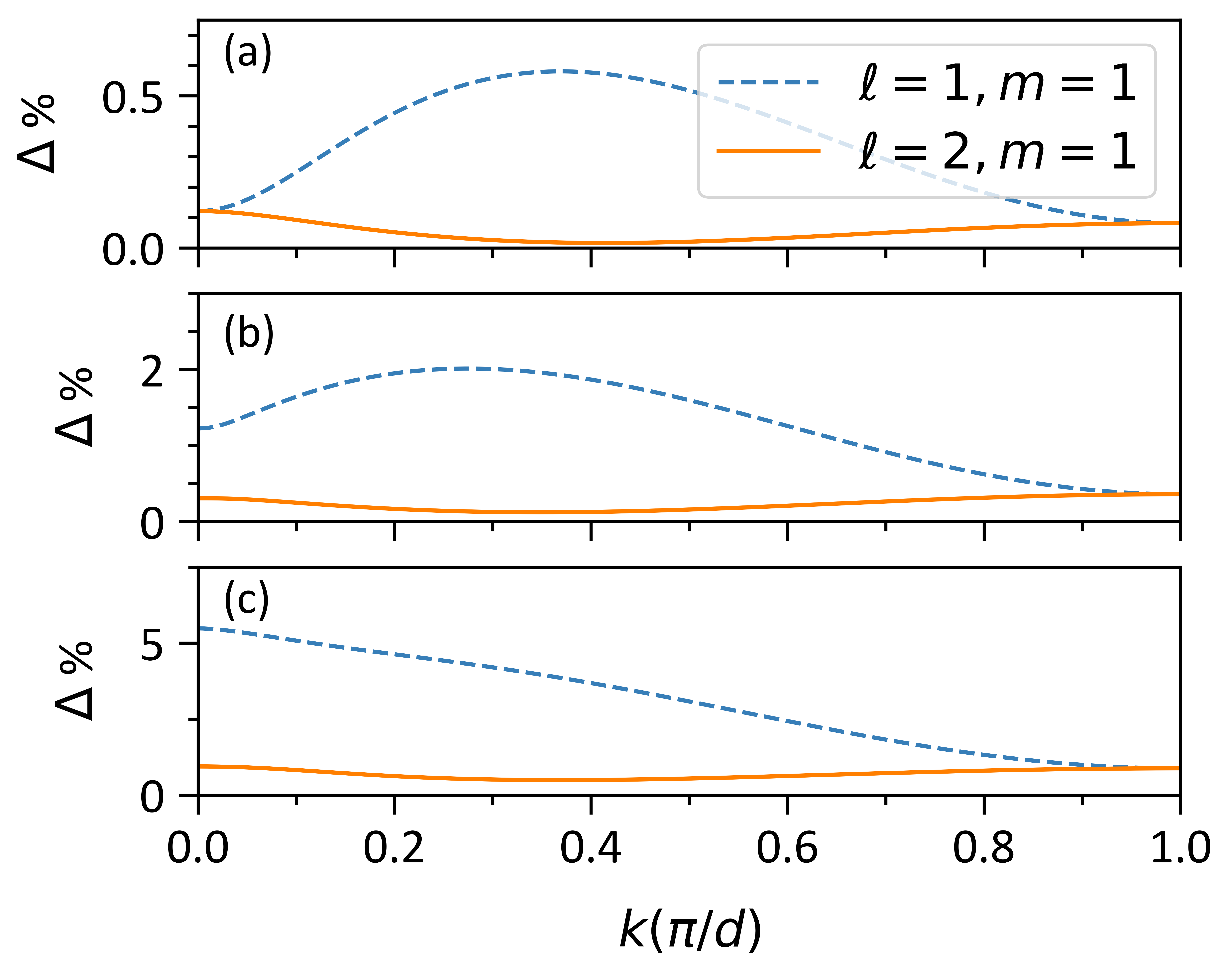} \caption{Relative energy differences $\Delta$ with regard to the converged
model $\ell_{max}=20$ across the Brillouin zone for (a) $d/R=3$,
(b) $d/R=2.6$ and (c) $d/R=2.4$ in the same order as in Fig.\ref{fig:bands}
and for both $m=0$ and $m=\pm1$. The expression for $\Delta$ is
$\Delta=\left|\left(\omega(k)-\omega_{\ell=20}(k)\right)/\omega_{\ell=20}(k)\right|$.}
\label{fig:relative} 
\end{figure}

The two highlighted bands in Fig. \ref{fig:bands} correspond to the
$m=0$ and $m=\pm1$ indexes, the latter being two times degenerate
owing to the symmetry of the chain. In the dipole model, the $m=0$
mode corresponds to dipoles aligned along the chain, while the $m=\pm1$
relates to transverse modes with dipoles orthogonal to the chain axis.
At the Brillouin zone center, the $m=0$-mode with all dipoles pointing
in the same direction is the fundamental excitation, as it minimizes
the dipolar tensor and can be viewed as a bonding mode. In contrast,
at the edge $k=\pi/d$, a N\'eel-like order for the $m=1$-mode is lower
in energy. This $\pi/d$ mode is referred to as an anti-bonding mode
in the hybridization picture proposed by Nordlander \emph{et al.}
\citep{Prodan}. Therefore, a band-crossing occurs as one goes from
$k=0$ to $\pi/d$, and this general observation does not depend on
the superlattice spacing. However, the dipole model remains quantitatively
valid only for $d/R\gtrsim3$ but at smaller inter-particle spacings
and finite $k$ this simple approach fails to provide an accurate
description. The blue dashed line in Fig. \ref{fig:relative} highlights
this fact: it represents the relative difference $\Delta(k)=\left|\left(\omega(k)-\omega_{\ell=20}(k)\right)/\omega_{\ell=20}(k)\right|$
with respect to the converged $\ell_{max}=20$-model. In addition,
this is further confirmed by the computation of the coefficient of
determination $\mathsf{R}^{2}$ 
\begin{equation}
\mathsf{R}^{2}=1-\frac{\sum_{k}(\omega(k)-\omega_{\ell=20}(k))^{2}}{\sum_{k}(\omega(k)-\overline{\omega_{\ell=20}})^{2}},
\end{equation}
which is given in Table \ref{tab:R2}. 
\begin{table}
\centering{}%
\begin{tabular}{lcccl}
\toprule 
 & \multicolumn{2}{c}{$\ell_{max}=1$} & \multicolumn{2}{c}{$\ell_{max}=2$}\tabularnewline
\midrule 
 & $m=0$  & $m=\pm1$  & $m=0$  & $m=\pm1$\tabularnewline
\midrule 
$d/R=3.0$  & 0.98  & 0.95  & 1.00  & 1.00\tabularnewline
$d/R=2.6$  & 0.88  & 0.71  & 1.00  & 0.99\tabularnewline
$d/R=2.4$  & 0.57  & 0.36  & 0.96  & 0.95\tabularnewline
\bottomrule
\end{tabular}\caption{Values of $\mathsf{R}^{2}$ coefficient between the $\ell=20$ values
and the truncated models values for the three lowest energy bands
of different superlattices geometry corresponding to the plasmonic
band diagrams of Fig. 2.}
\label{tab:R2} 
\end{table}

In contrast, the dipole+quadrupole model, depicted by the blue crosses
in Fig. \ref{fig:bands} remains quantitatively
valid down to $d/R=2.4$. As one can see in Fig. \ref{fig:relative},
truncating at $\ell_{max}=2$ leads to a relative energy difference
that remains below $3\%$ across the whole $k$-space, even for dense
assemblies and an $\mathsf{R}^{2}$, which is larger than $95\%$,
comparable to that of the dipole model at $d/R=3$.
Taking into account the higher-order $\ell=3$ octupolar modes corrects
the model prediction only slightly as can be seen in Fig. \ref{fig:bands},
with $\mathsf{R}^{2}$ values of 0.99 at $d/R=2.4$ for both bands.
In principle, the exact solution implies a Hamiltonian diagonalization
with $\ell_{max}\to\infty$, it is therefore a matter of taste to
truncate at an arbitrary mode once a convergence criterion is satisfactory.
Of course, involving the octupoles implies a better quantitative match,
but limiting ourselves to a dipole+quadrupole truncation has the advantage
of providing a good physical description of the main low-lying excitations
(within 5\% quantitatively speaking). Besides, this approach allows
one to obtain the tractable semi-analytical dispersion of Eq. (\ref{eqn:wk}),
and provides a clear framework to understand the role of higher-order
multipoles in plasmonic bands, as we shall see in the next Section.

As for the calculation of the plasmonic eigenvalues
at a given $\bm{k}$, the dimensions of the Hamiltonian to be solved
increase as $\sum_{\ell=1}^{\ell_{max}}(2\ell+1)=\ell_{max}^{2}+2\ell_{max}$.
This gives Hamiltonian sizes of 3, 8, and 440 for the dipolar model,
the quadrupolar model, and the model with $\ell_{max}=20$ respectively.
From an experimental point of view, the main interest relies in the
high symmetry $k=0$ point. To quantify the energy difference between
the dipole model prediction, and the dipole+quadrupole model at this
point, let us consider silver nanoparticles with $\hbar\omega_{p}=9.04{\rm \ eV}$.
The dipole model's predictions differ from those of the converged
model $\ell_{max}=20$ by $\Delta E_{m=0}(k=0)=0.09{\rm \ eV}$ and
$\Delta E_{m=1}(k=0)=0.29{\rm \ eV}$ while for the dipole+quadrupole
model we have $\Delta E_{m=0}(k=0)=0.09{\rm \ eV}$ and $\Delta E_{m=1}(k=0)=0.05{\rm \ eV}$.

The high precision that is obtained from the simple truncation at
$\ell_{max}=2$ comes from the very structure of the Hamiltonian of
Eqs. (\ref{eqn:Hpl}) and (\ref{eqn:Qlmk}). Indeed, as was previously
emphasized, only terms coupling the same $m$ remain; this gives rise
to a sparse block-diagonal matrix, which combines to the decay in
$d^{-\left(\ell+\ell'+1\right)}$ to make the higher-order modes contribute
very little. To sum up, these results confirm that the dispersion
of Eq. (\ref{eqn:wk}) is sufficient to describe low-energy physics
in plasmonic nanoparticle chains.

\subsection{Eigenstate analysis: role of the quadrupoles}

To determine more precisely the role of the quadrupoles in the system,
let us inspect the eigenstates of the low-lying plasmonic bands. These
eigenstates can be written as linear combinations of the single-particle
multipoles 
\begin{equation}
\left|\psi_{m}(k)\right\rangle =a_{1m}(k)\left|\phi_{1m}\right\rangle +a_{2m}(k)\left|\phi_{2m}\right\rangle ,
\end{equation}
where the expressions of the coefficients $a_{1m}(k)$ and $a_{2m}(k)$
depend on the ordering of the subspaces of dipoles $\omega_{1k}^{2}$
and quadrupoles $\omega_{2k}^{2}$ and their difference $\Delta_{m}(k)=\omega_{1}^{2}+\omega_{p}^{2}Q_{k11m}-\omega_{2}-\omega_{p}^{2}Q_{k22m}$,
appearing in the square root of Eq. (\ref{eqn:wk}). This leads to
the following rewriting 
\begin{equation}
\left\{ \begin{array}{lll}
\left|\psi_{m,>}(k)\right\rangle  & = & -\sin\frac{\theta_{m}(k)}{2}\left|\phi_{1m}\right\rangle +\cos\frac{\theta_{m}(k)}{2}\left|\phi_{2m}\right\rangle \\
\\\left|\psi_{m,<}(k)\right\rangle  & = & \cos\frac{\theta_{m}(k)}{2}\left|\phi_{1m}\right\rangle -\sin\frac{\theta_{m}(k)}{2}\left|\phi_{2m}\right\rangle 
\end{array}\right.\label{eq:Psi}
\end{equation}
where $\left|\psi_{m,>}(k)\right\rangle $ stands for the case $\Delta_{m}(k)>0$,
whilst $\left|\psi_{m,<}(k)\right\rangle $ holds for $\Delta_{m}(k)<0$,
and the phase $\theta_{m}(k)$ is given by $\theta_{m}(k)=-\tan^{-1}\left|\frac{2Q_{k12m}}{\Delta_{m}(k)}\right|$.

\begin{figure}
\centering{}\includegraphics[width=0.5\textwidth]{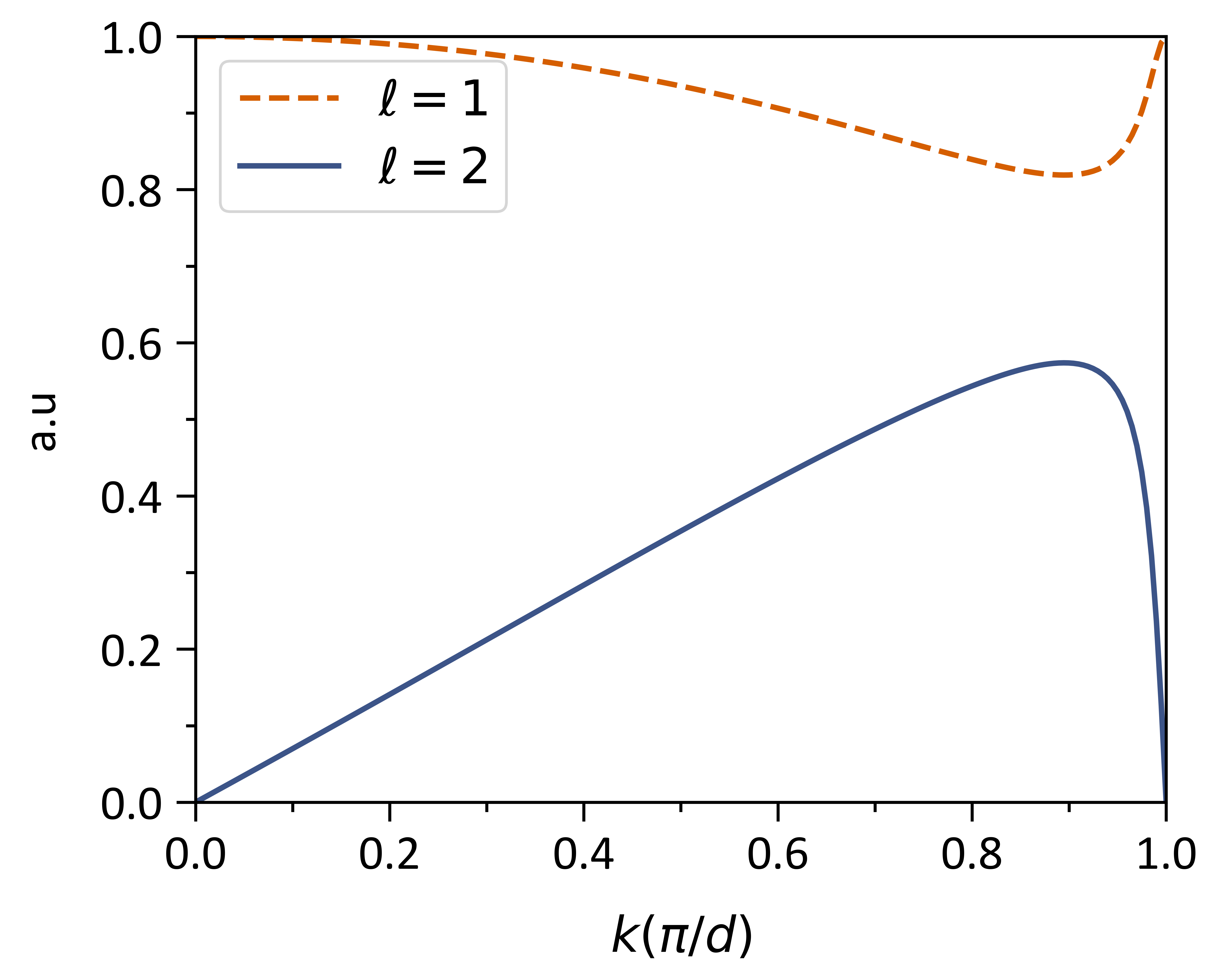}
\includegraphics[width=0.5\textwidth]{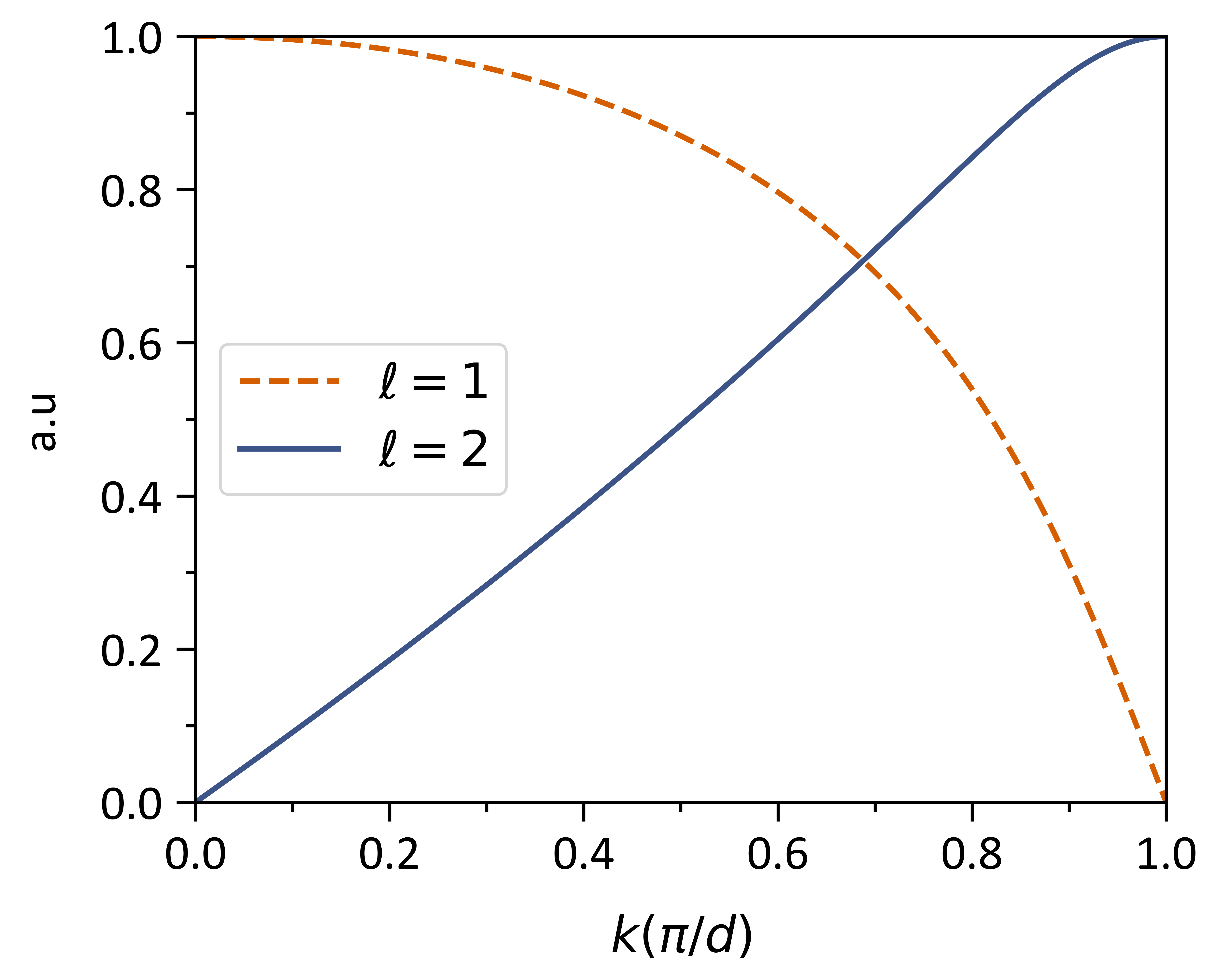}
\includegraphics[width=0.5\textwidth]{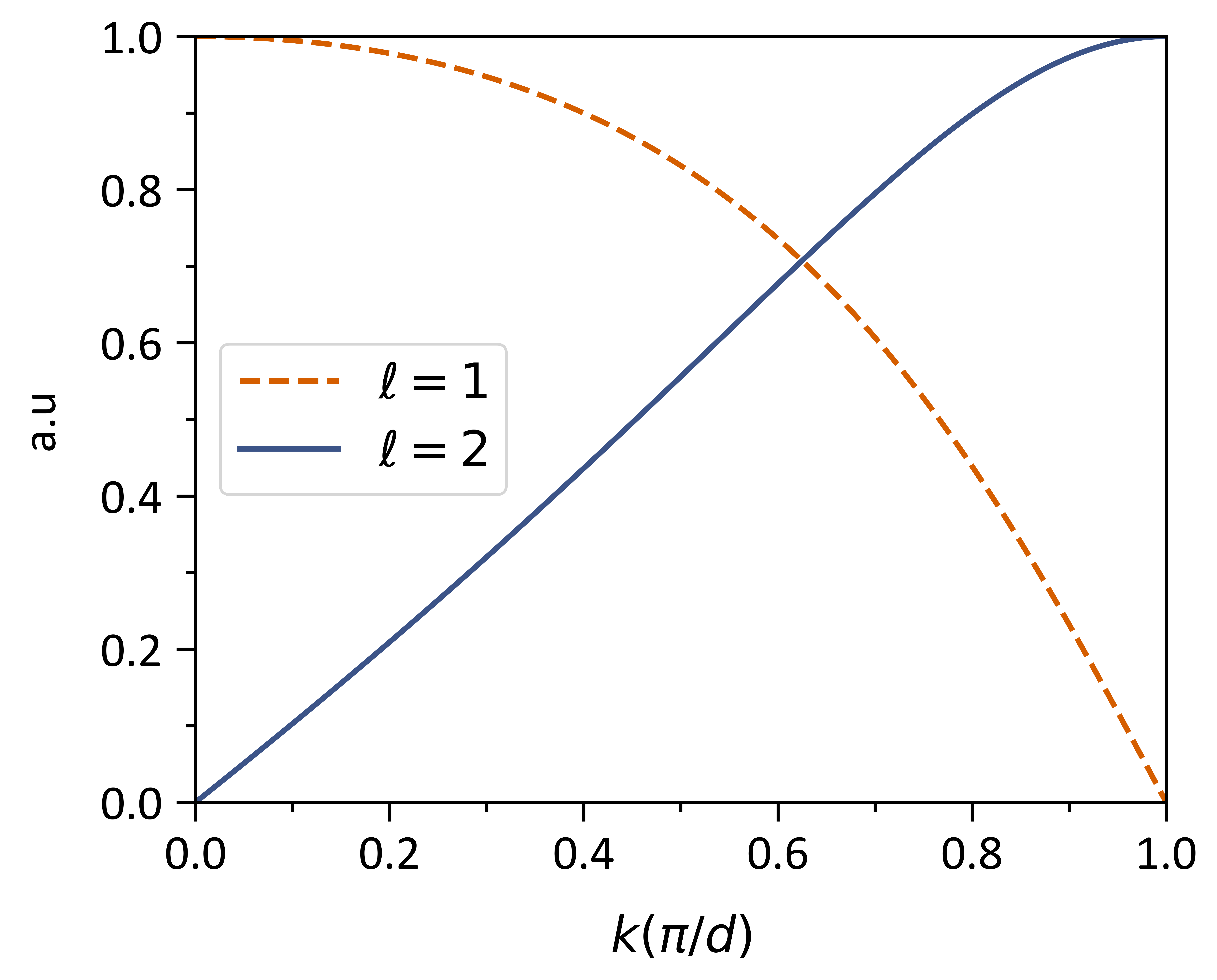}
\caption{(a) Components of the eigenfunctions associated with the band dispersion
(\ref{eqn:wk}) for the $m=0$ band of a chain of spherical nanoparticles
of radius $R=10$nm with three different lattice spacing (a) $d=30nm$,
(b) $d=26$nm and (c) $d=24$nm. Red dashed lines are the dipole component
$a_{\ell=1,m=0}(k)$, and the blue lines are the quadrupolar components
$a_{\ell=2,m=0}(k)$}
\label{fig:compm0} 
\end{figure}

\begin{figure}
\centering{}\includegraphics[width=0.5\textwidth]{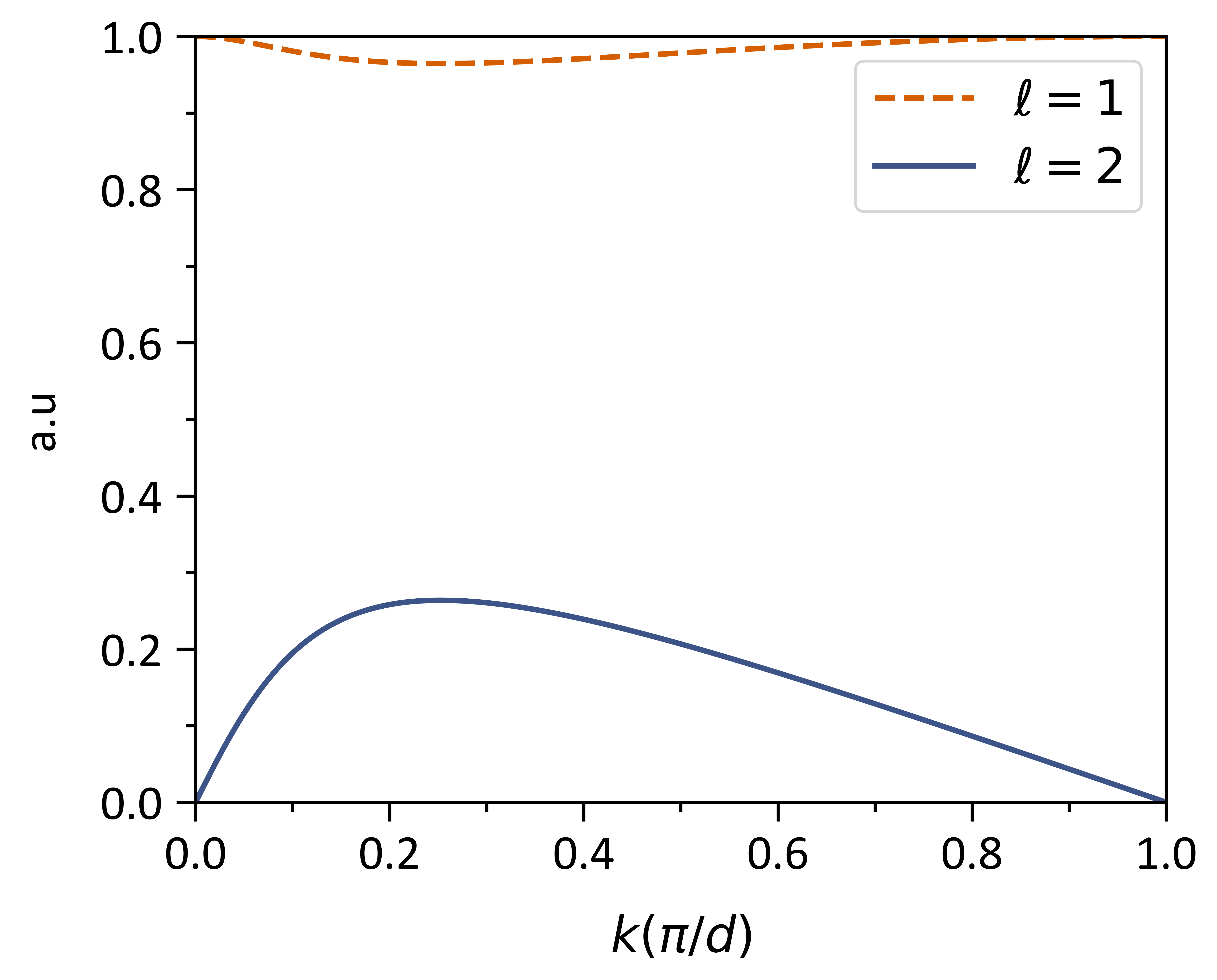}
\includegraphics[width=0.5\textwidth]{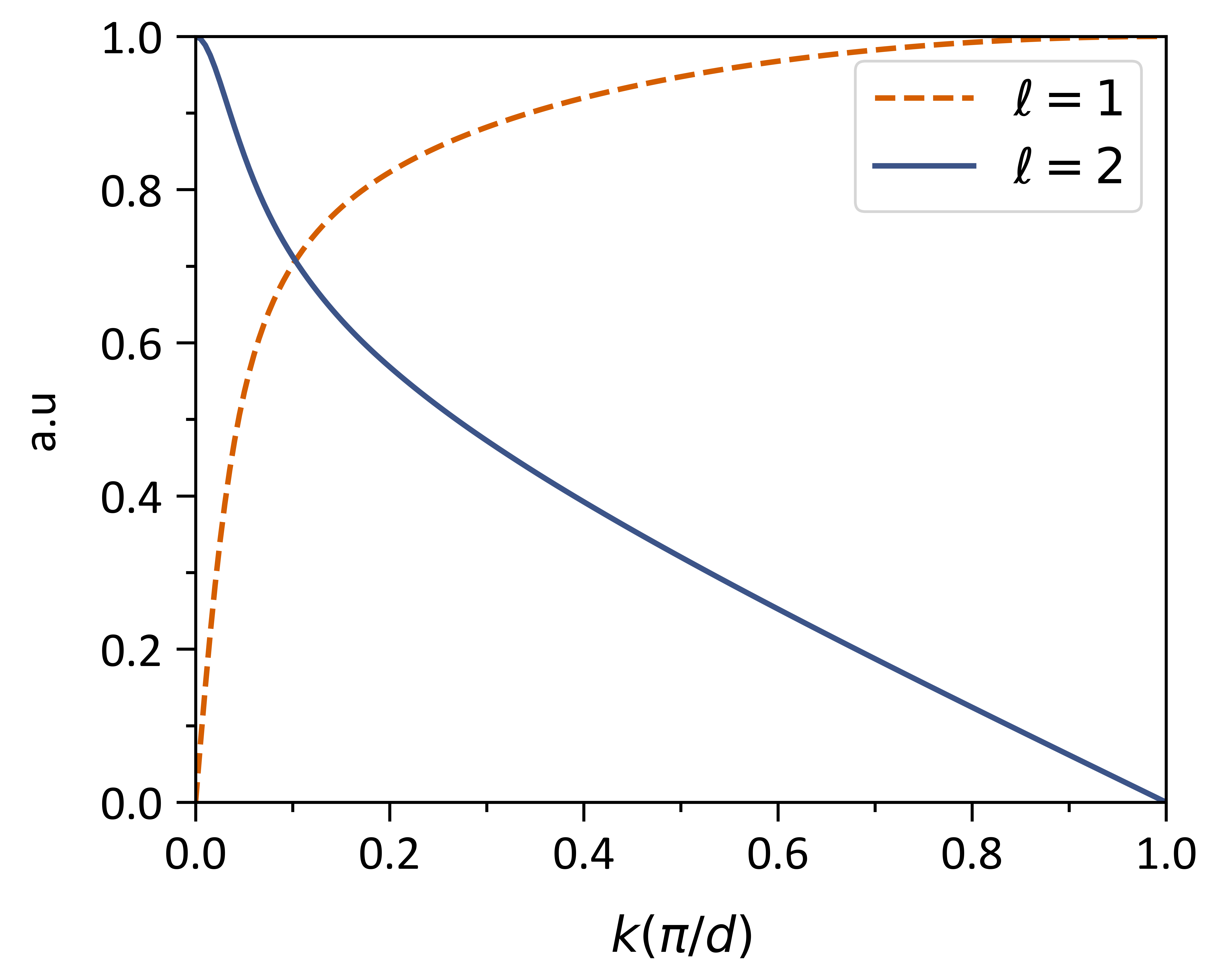}
\includegraphics[width=0.5\textwidth]{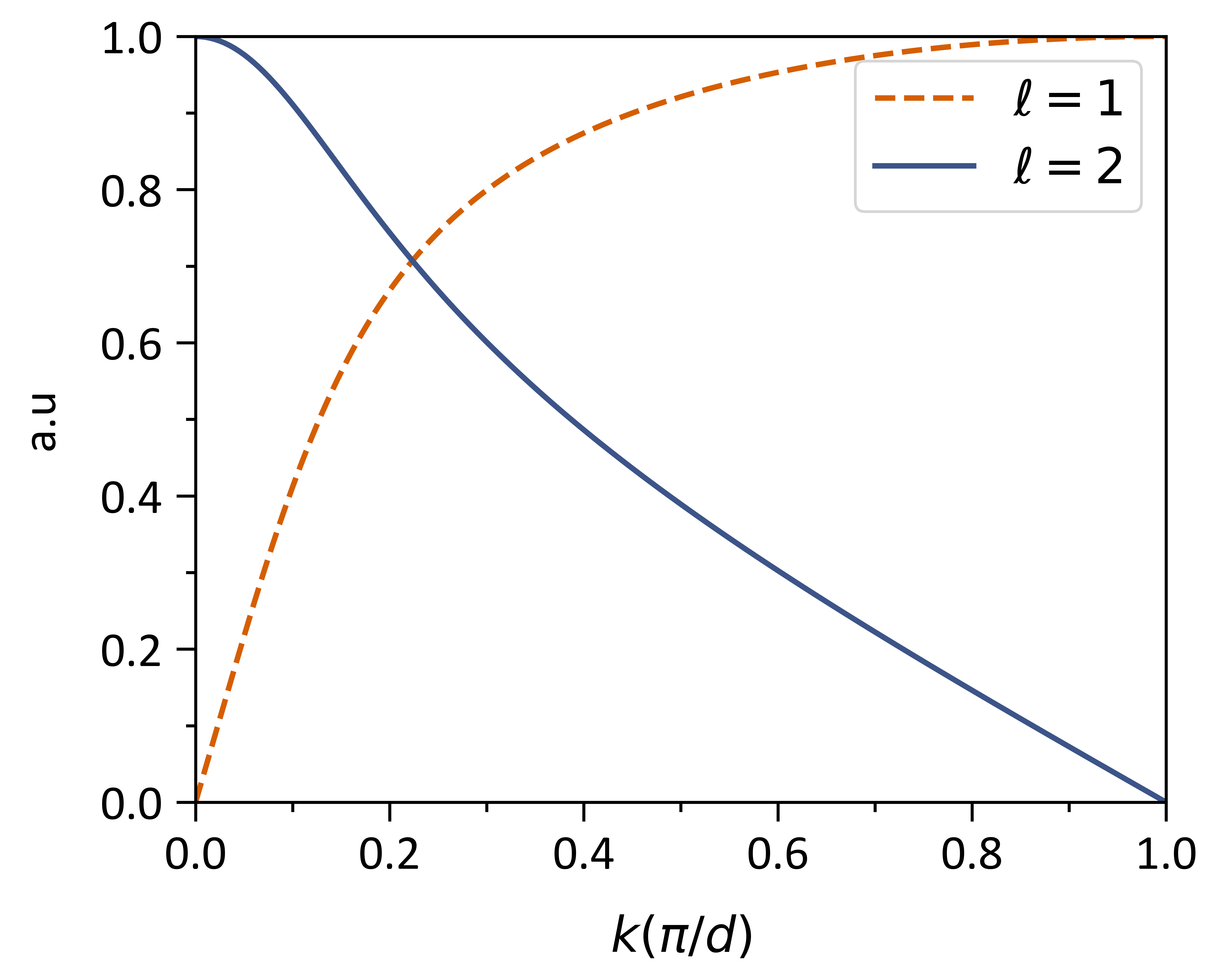}
\caption{(a) Components of the eigenfunctions associated with the band dispersion
(\ref{eqn:wk}) for the $m=\pm1$ band of a chain of spherical nanoparticles
of radius $R=10$nm with three different lattice spacing (a) $d=30nm$,
(b) $d=26$nm and (c) $d=24$nm. Red dashed lines are the dipole component
$a_{\ell=1,m=1}(k)$, and the blue lines are the quadrupolar components
$a_{\ell=2,m=1}(k)$}
\label{fig:compm1} 
\end{figure}

Fig. \ref{fig:omega2} ($d/R=2.4$) shows that the sign of $\Delta_{m}(k)$
might change along the Brillouin-zone, implying a crossing of dipolar
and quadrupolar subspaces. Therefore, as seen in Eqs. (\ref{eq:Psi}),
the main component of the mode changes, switching from dipoles to
quadrupoles or \emph{vice versa}. If the lattice spacing is large
enough ($d/R\gtrsim3$), this crossing does not occur since quadrupoles
are too high in energy and almost decoupled from dipoles. In contrast,
when the particles are close enough, the admixture of the modes can
no longer be neglected as their energies are comparable. This implies
that even though one focuses on low-energy physics, quadrupoles must
be considered in the model since they become a non-negligible component
of the bands. To highlight this fact, we plot the $a_{1m}(k)$ and
$a_{2m}(k)$ coefficients for the three different lattice spacings
$d/R=3$, $d/R=2.6$ and $d/R=2.4$, and the two different bands $m=0$
and $m=\pm1$, in Figs. \ref{fig:compm0} and \ref{fig:compm1}.

For $d/R=3$, depicted at the top of Figs. \ref{fig:compm0} and \ref{fig:compm1},
as expected, the dipolar component largely dominates the wave-function
of Eq. (\ref{eq:Psi}) across the whole Brillouin zone. In the case
of the $m=0$ band, and for both $d/R=2.6$ and $d/R=2.4$, the wave-function
is purely dipolar at $k=0$, then it keeps a major dipolar contribution
for small wave-vectors. At larger ones, a crossing occurs, and a pure
quadrupolar state ends to take place at $k=\pi/d$. At high-symmetry
points, quadrupoles and dipoles do not couple, which is in line with
the results of Barros et al. \citep{Barros}, they only mix at finite
wave-vectors and eventually cross. Even though the crossing wave-vector
gets smaller as the lattice spacing is reduced, it remains in the
order of $k_{{\rm cross}}\sim0.6\pi/d$, which means that this would
hardly couple to light.

More interestingly, for the $m=\pm1$ bands of Fig. \ref{fig:compm1},
the nature of the state at $k=0$ changes as the lattice spacing is
reduced: From a pure dipolar state for $d/R=3$, it becomes purely
quadrupolar for $d/R=2.6$ and $d/R=2.4$. As $k$ is increased, the
dipolar and quadrupolar components mix, up to the crossing point $k_{{\rm cross}}\simeq\left(0.1-0.2\right)\pi/d$
from which the dipolar state becomes dominant. In contrast to the
previous $m=0$ case, the crossing point is close to the Brillouin
zone center and, therefore, accessible to light-scattering.

As a result, it should be experimentally possible to measure the relative
contribution of each component, dipole and quadrupole, and to see
the physical nature of the state change if the assembly is more or
less dense. However, this work of coupling to light and a photonic
bath implies a modification of the eigenstates and is left for further
investigation.

\section{Conclusion and perspectives}

Having developed a quasistatic Hamiltonian (\ref{eqn:Hpl}) from the
plasmonic overlap of classical and the quantum models for plasmonic
arrays, we have shown that dipolar and quadrupolar modes represent
the major contribution of the low-energy physics and can be decoupled
from higher-order modes. Consequently, we proposed a minimal model
that treats dipoles and quadrupoles on the same footing and allows
an accurate description of the two first plasmonic excitations across
the whole Brillouin zone of a chain of nanoparticles. Further inspection
of the nature of the modes underlined the importance of quadrupoles
in dense particle arrays, solely due to plasmon-plasmon interactions.
For this reason, in future work considering also the coupling with
light, the quadrupole will have to be accounted for regardless of
how little they couple to light. We pointed out that even though
quadrupoles do not couple to dipoles at high symmetry points ($k=0$
for the chain), they do at others, rendering them essential to investigate
plasmon-polaritons in nanoparticle assemblies. The coupling of the
present Hamiltonian to a photonic bath $H_{\gamma}$ that would allow
taking account of retardation effects and computing physical observables
is left for further investigation. Finally, the second quantized formalism
used here makes our Hamiltonian relevant for coupling to other systems,
such as molecules or quantum dots, that couple to the highly localized
fields of quadrupoles.

\section*{Acknowledgments}

This work is supported by the Labex SOLSTICE (ANR-10-LABX-0022)

\section{References}

 \bibliographystyle{unsrtnat}
\bibliography{biblio}

\end{document}